\pdfoutput=1

\documentclass[11pt]{journal}

\usepackage{fancyhdr}
\usepackage{times}
\usepackage{epsfig,subfigure,amssymb,amsmath,makeidx}
\usepackage{cite, chapterbib}
\usepackage{makeidx}
\usepackage{procedure}
\usepackage{algorithm}
\usepackage{algorithmic}
\usepackage{enumerate}
\usepackage{url}
\usepackage{floatflt}
\usepackage{wrapfig}
\usepackage{boxedminipage}
\usepackage{backref}

\usepackage{graphicx}

\makeindex
\newcommand{\bea}{\begin{eqnarray*}}
\newcommand{\eea}{\end{eqnarray*}}

\newcommand{\BE}{\begin{equation}}
\newcommand{\EE}{\end{equation}}
\newcommand{\BEA}{\begin{eqnarray}}
\newcommand{\EEA}{\end{eqnarray}}

\newcommand{\barray}{\begin{array}{ll}}
\newcommand{\earray}{\end{array}}

\newcounter{neweqn}

\newcommand{\beq}[1]{\begin{equation} \refstepcounter{neweqn} \label{#1}}
\newcommand{\eeq}{\end{equation}}
\newcommand{\bed}{\begin{displaymath}}
\newcommand{\eed}{\end{displaymath}}

\def\one{{\hbox{1{\kern -0.35em}1}}}

\def\({\left(}
\def\){\right)}

\newcommand{\ben}{\begin{eqnarray*}}
\newcommand{\een}{\end{eqnarray*}}

\def\xd#1 {\fbox {\footnote {\ }}\ \footnotetext {xd comment: #1}}
\def\rbt#1 {\fbox {\footnote {\ }}\ \footnotetext {rbt comment: #1}}

\newtheorem{theorem}{Theorem}[section]

\newcommand{\qed}{\nobreak \ifvmode \relax \else
      \ifdim\lastskip<1.5em \hskip-\lastskip
      \hskip1.5em plus0em minus0.5em \fi \nobreak
      \vrule height0.75em width0.5em depth0.25em\fi}

\begin{document}
   
		  	\thispagestyle{empty}
				\vspace{6in}


\clearpage\setcounter{page}{1} 

 	  \clearpage\setcounter{page}{1}

\begin{center}
{\huge{\textbf{The Foundation of Big Data: Experiments, Formulation, and Applications}}}

\end{center}

\begin{center}
{\large{\textbf{Robert C. Qiu }}}\footnote{The authors are with Department of Electrical and Computer Engineering, Tennessee Technological University, Cookeville, Tennessee 38505. Email: rqiu@tntech.edu or rqiu@ieee.org. Website: http://iweb.tntech.edu/rqiu/index.htm. }
\end{center}

\section{INTRODUCTION}
	The central theme of this talk~\cite{Qiu2014BigData} is to promote the non-asymptotic statistical viewpoint in the context of massive datasets.  The classical viewpoint breaks down when the data size becomes large.

\section{Non-Asymptotic Viewpoint}

In~\cite{lin2012generalized}, the author's group has shown that the classical GLRT is not optimal for large data lengths. This discovery has triggered a series of books~\cite{QiuBook2012Cognitive,Qiu_WicksBook2013,QiuAntonik2014Wiley}. 

Very recently\footnote{The author thanks Dr. Husheng Li at University of Tennessee, at Knoxville for his pointing out this line of work in September 2014.},  the work of Polyanskiy, Poor and Verdu (2010)~\cite{polyanskiy2010channel} has come to our attention. In finite-blocklength coding, Polyanskiy, Poor and Verdu  proves the following result 
\begin{equation}
\label{eq:polyanskiy2010channel}
	R\left( {\varepsilon ,n} \right) = C - \sqrt {\frac{V}{n}} {Q^{ - 1}}\left( \varepsilon  \right) + O\left( {\frac{1}{{\sqrt n }}} \right),
\end{equation}
where $\varepsilon$  is the error probability, $n$ is the coding length, $V$ is the dispersion and $Q(x)$ is the standard error function.  This result says that Shannon's result on channel capacity $C$ is only the asymptotic term. His result is not optimal for finite blocklengths.  As a matter of fact, the second order term $\sqrt {\frac{V}{n}} {Q^{ - 1}}\left( \varepsilon  \right)$ is explicitly given in~\eqref{eq:polyanskiy2010channel}.

The spirits of two above works are identical. The classical results are optimal only in the asymptotic limit $n\to \infty.$  In the age of Big Data, the size of $n$ is large, but FINITE! The large but finite $n$ asks for a new paradigm shift in analytical tools. 

\section{Work at TTU Since 2010 }
High scalability is the most important requirement for big data analysis. Size is the only thing that matters. It is natural to model massive datasets by using large random matrices. In the asymptotic regimes, the sizes of random matrices are assumed to approach the infinity. For example, for a random matrix  $\bf X$   of size  $m\times n,$ we assume the asymptotic regime:  \[m \to \infty ,n \to \infty {\text{  but  }}m/n \to c \in \left( {0,\infty } \right).\] The work~\cite{QiuBook2012Cognitive} give a survey of the work in this direction and study the connection with cognitive radio network.

The non-asymptotic regime is defined as: $m$ and $n$ are large, but finite.  In~\cite{Qiu_WicksBook2013}, we use concentration inequalities as the mathematical foundation. The book goes to study compressed sensing, low rank matrix recovery, estimation, detection and database-friendly computing.  Large random matrices can unify many big data problems and form the foundation for big data analysis.  In~\cite{QiuAntonik2014Wiley}, we build upon this key observation. There we go on to study non-Hermitian random matrices using free probability theory.  The three books promote the framework of using large random matrices (Hermitian or non-Hermitian) to represent massive datasets.

\section{Data Collection and Modeling}
\subsection{Data Collection}
\begin{figure}
	 \centering
		 \includegraphics[scale=0.5]{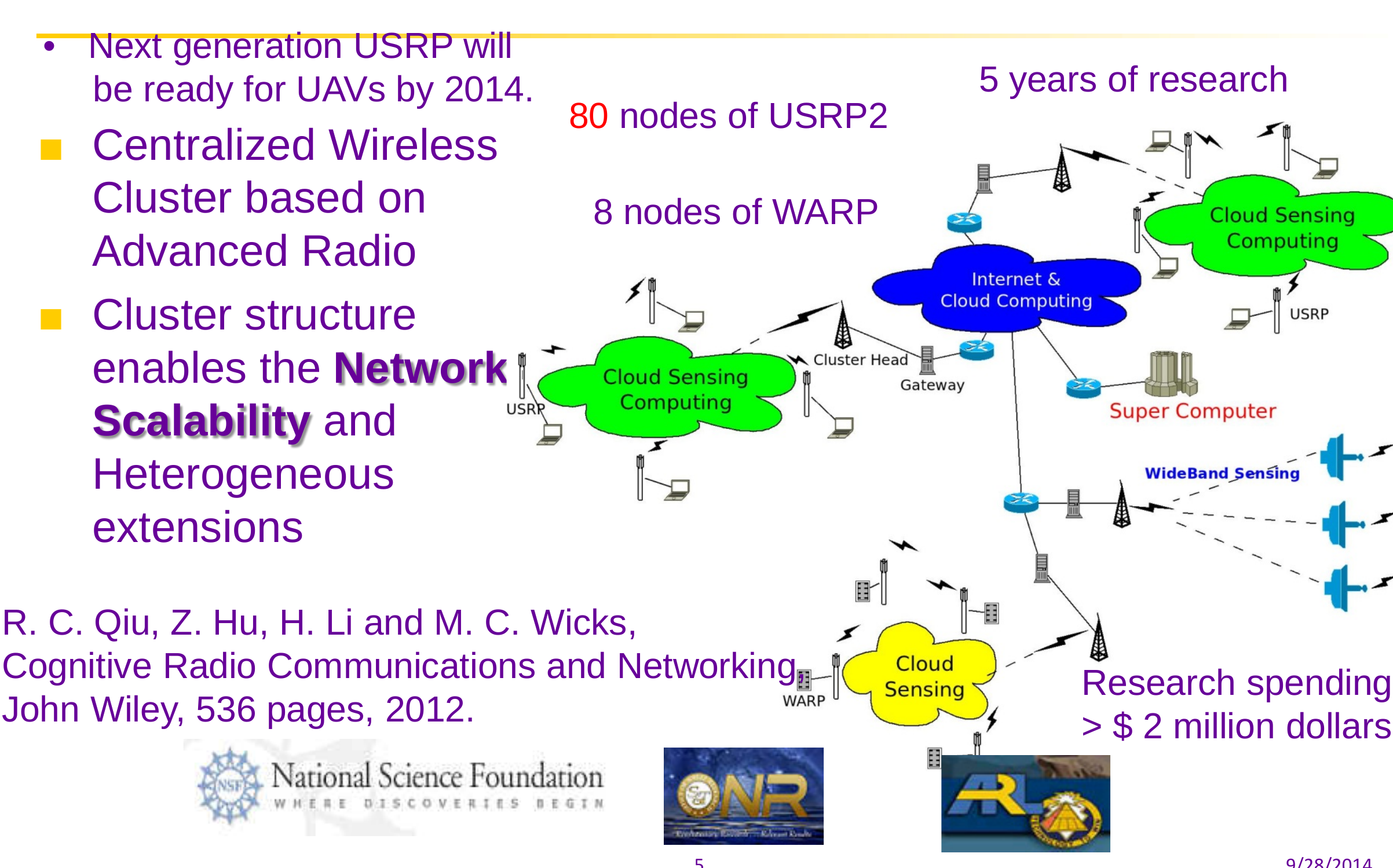}
	 \caption{A Large-Scale Advanced Radio Network Testbed.}
	 \label{fig:LargeScaleTestbed}
 \end{figure}

Figure~\ref{fig:LargeScaleTestbed} shows a 80-node network made of software-defined radios (USRP). The potential research topics include: time-varying topology, UAVs, random arrays, anti-jamming, network security and privacy of data.

The central message here is to show the big data problem arose from the natural engineering environment.  Google downloads all webpages from the Internet, and saves them for data mining.  Big Data at least traces back to its practice. In the same spirit, can we save and makes sense of  ''all the data'' in wireless networks? The answer is a big YES. TTU has saved all the radio waveforms (modulated waveforms) and attempted to make sense of these massive datasets. Real-time, streaming data of massive size is the hallmark of this work.

\begin{figure}
	\centering
		\includegraphics[scale=0.8]{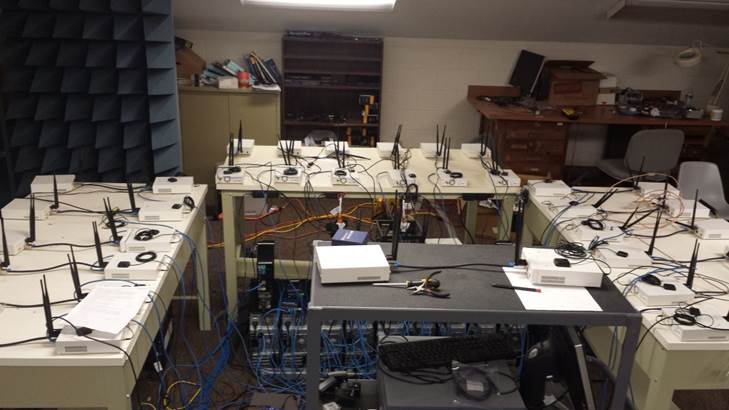}
	\caption{ The photo of the 80-node testbed at TTU.}
	\label{fig:photoUSRPnetwork}
\end{figure}

\begin{figure}
	\centering
		\includegraphics[scale=0.6]{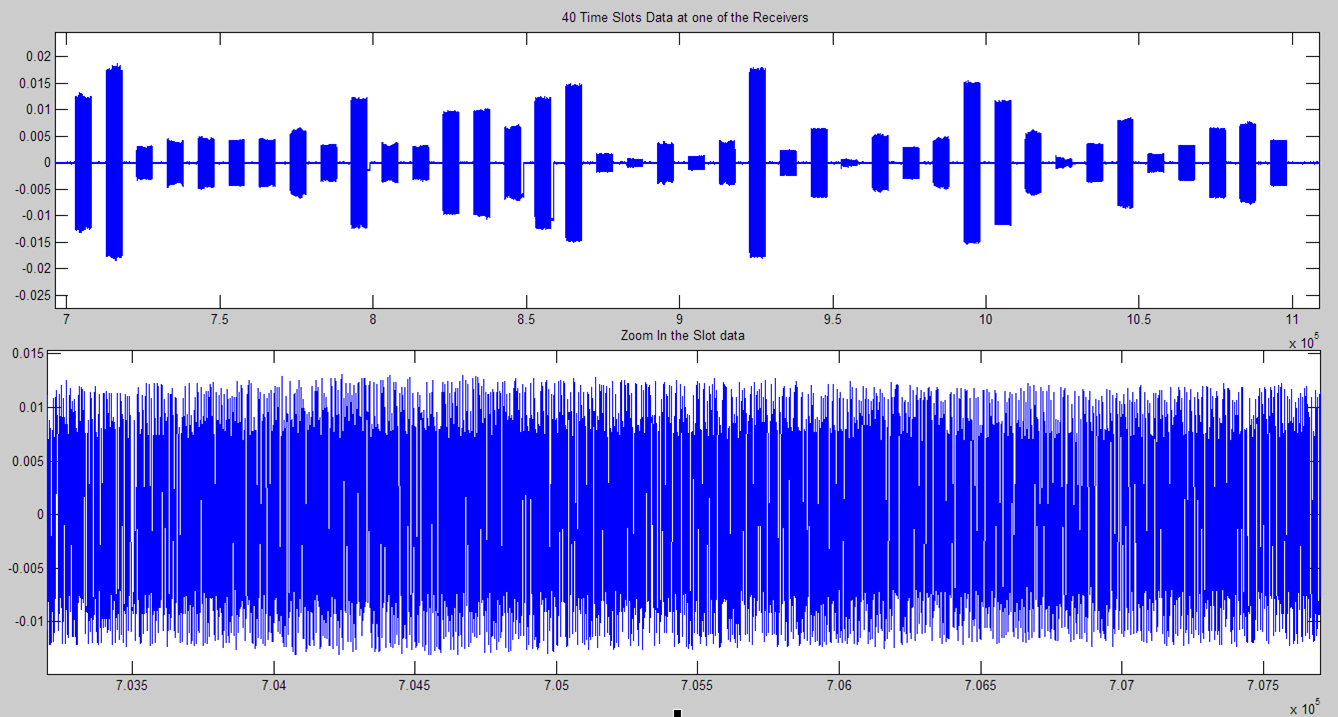}
	\caption{Waveforms of  soft-defined radios. The above represents 40 time slots of data recorded at one USPR receiver; the bottom is zoomed in for each time slot. The sample rate is 20 Msps.}
	\label{fig:WaveformsUSRP}
\end{figure}

 Figure~\ref{fig:photoUSRPnetwork} shows the real deployment of TTU's cognitive radio network. We have 80-node USRPs. Massive MIMO can be studied in this testbed. Figure~\ref{fig:WaveformsUSRP} illustrates waveforms of 80 soft-defined radios. One asks a basic question: How do we represent these massive datasets? In the next subsection, we propose to use large random matrices for this purpose.

\subsection{Large Random Matrices}

In~\cite{zhang2014data,li2014modeling}, we model the massive datasets using large random matrices.

\subsection{Empirical Spectrum Density}

\begin{figure}
	\centering
		\includegraphics[scale=0.6]{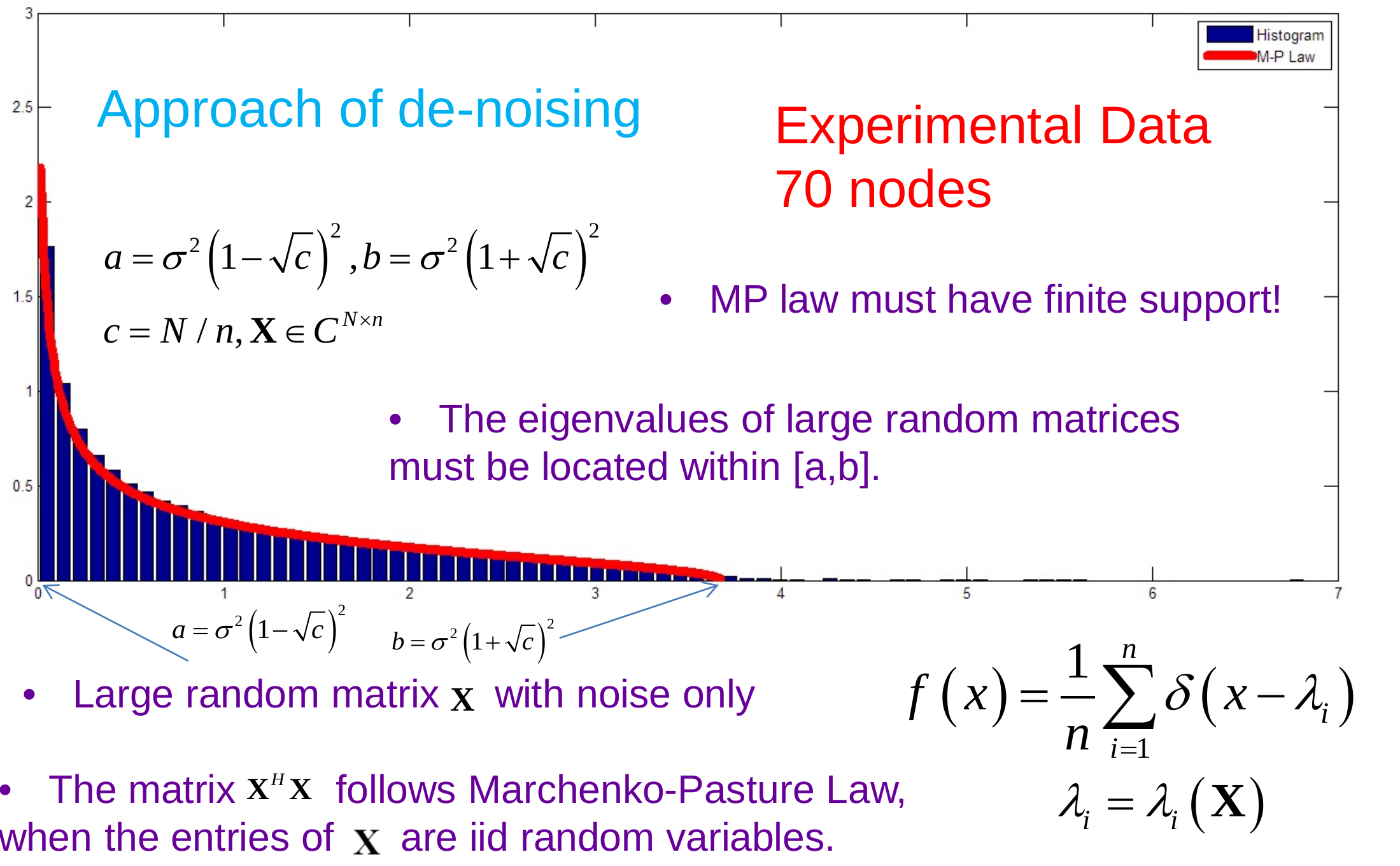}
	\caption{Empirical Spectral Density: Noise Only}
	\label{fig:EmpiricalsESPNoiseonly}
\end{figure}

Figure~\ref{fig:EmpiricalsESPNoiseonly} shows empirical spectral density: noise only. 

\begin{figure}
	\centering
		\includegraphics[scale=0.6]{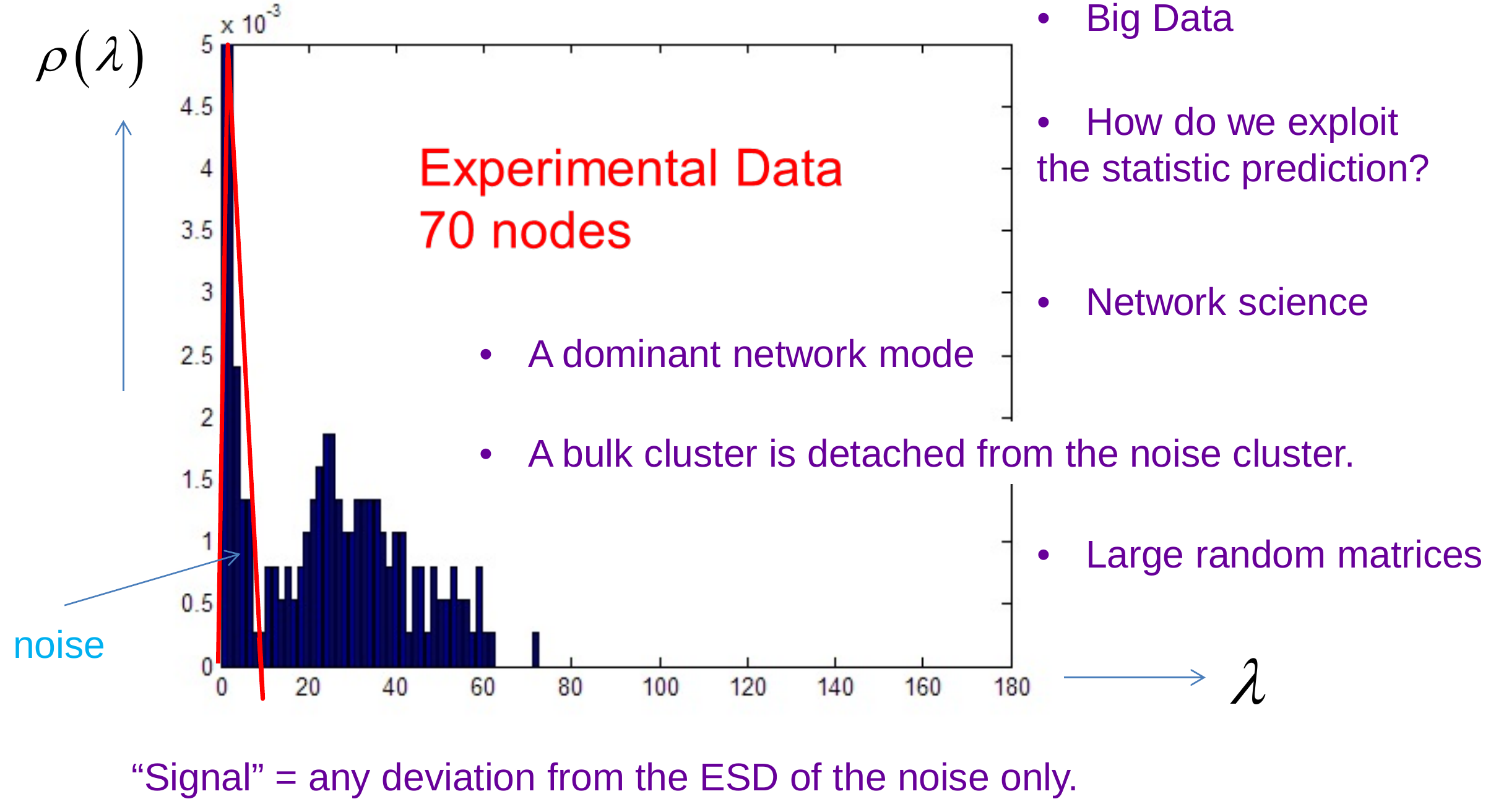}
	\caption{Empirical Spectral Density: Signal plus Noise}
	\label{fig:ESDsignalplusnoise}
\end{figure}

Figure~\ref{fig:ESDsignalplusnoise} shows empirical spectral density: signal plus noise only. Any deviation from ''noise only'' case will indicate the presence of ''some signal''. We observe a bulk cluster that is detached from the noise cluster. Note the noise cluster follows the Marchenko-Pasture law, as shown in Figure~\ref{fig:EmpiricalsESPNoiseonly}.

\subsection{Product of Non-Hermitian Random Matrices}
\begin{figure}
	\centering
		\includegraphics[scale=0.6]{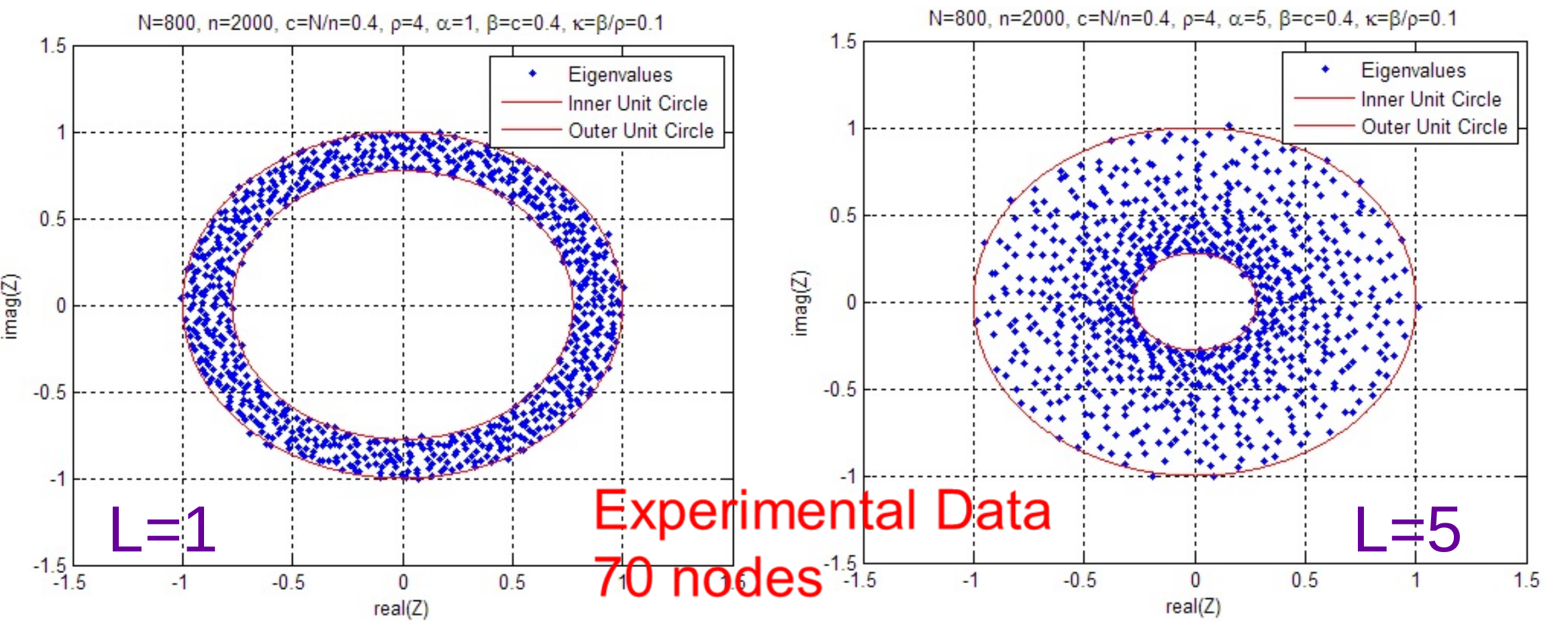}
	\caption{Product of Non-Hermitian Random Matrices: Noise Only}
	\label{fig:RingLawNoisely70nodes}
\end{figure}

Figure~\ref{fig:RingLawNoisely70nodes} shows the product of non-Hermitian random matrices: noise only. The product of non-Hermitian random matrices can be studied by the new free probability theory. The ring law is so universal for a basic building block, a rectangular random matrix. 

\begin{figure}
	\centering
		\includegraphics[scale=0.6]{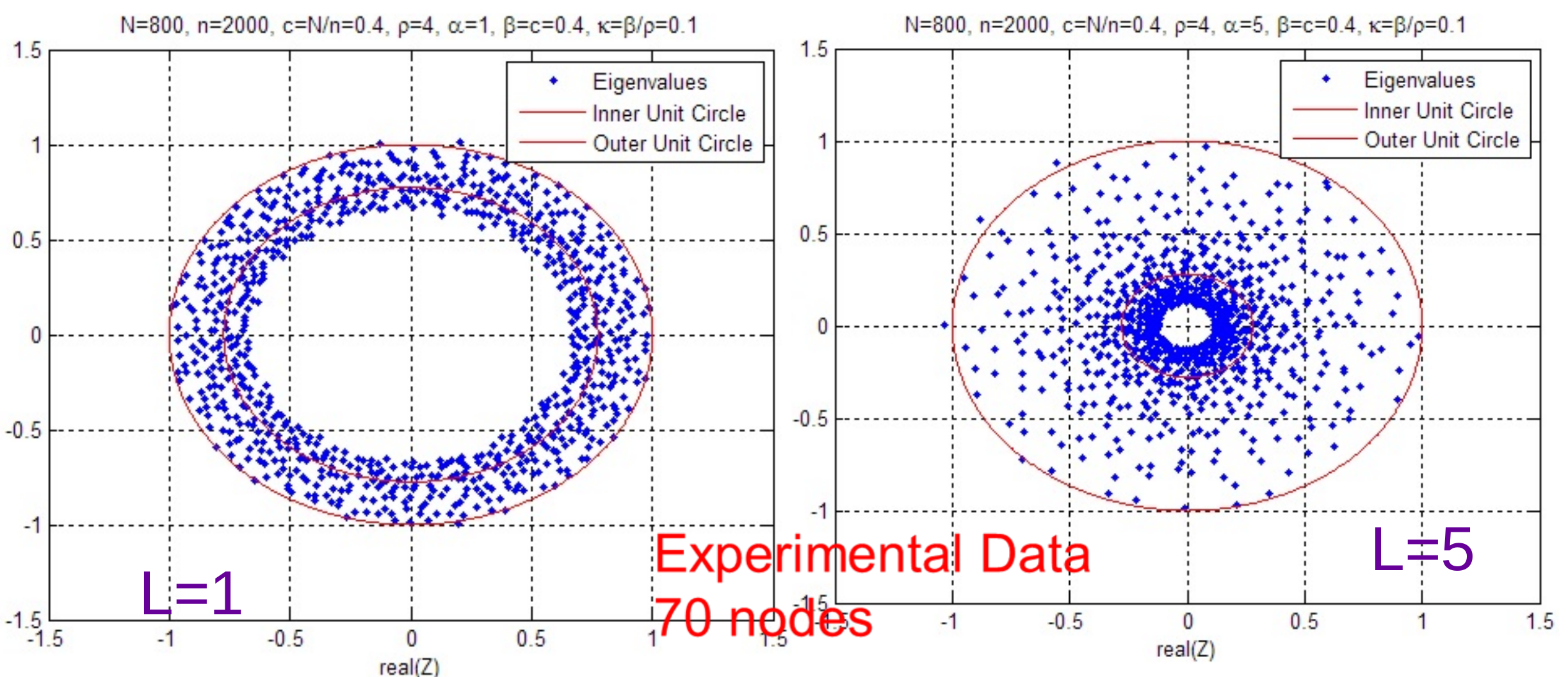}
	\caption{Product of Non-Hermitian Random Matrices: Signal plus Noise}
	\label{fig:RingLawSignalNoisely70nodes}
\end{figure}

Figure~\ref{fig:RingLawSignalNoisely70nodes} shows the product of non-Hermitian random matrices: signal plus noise. Compared with the previous case of  ''Noise only'', correlation of the signal reduces the inner radius of the ring law.  We also observed this using data in other fields.

\begin{figure}
	\centering
		\includegraphics[scale=0.6]{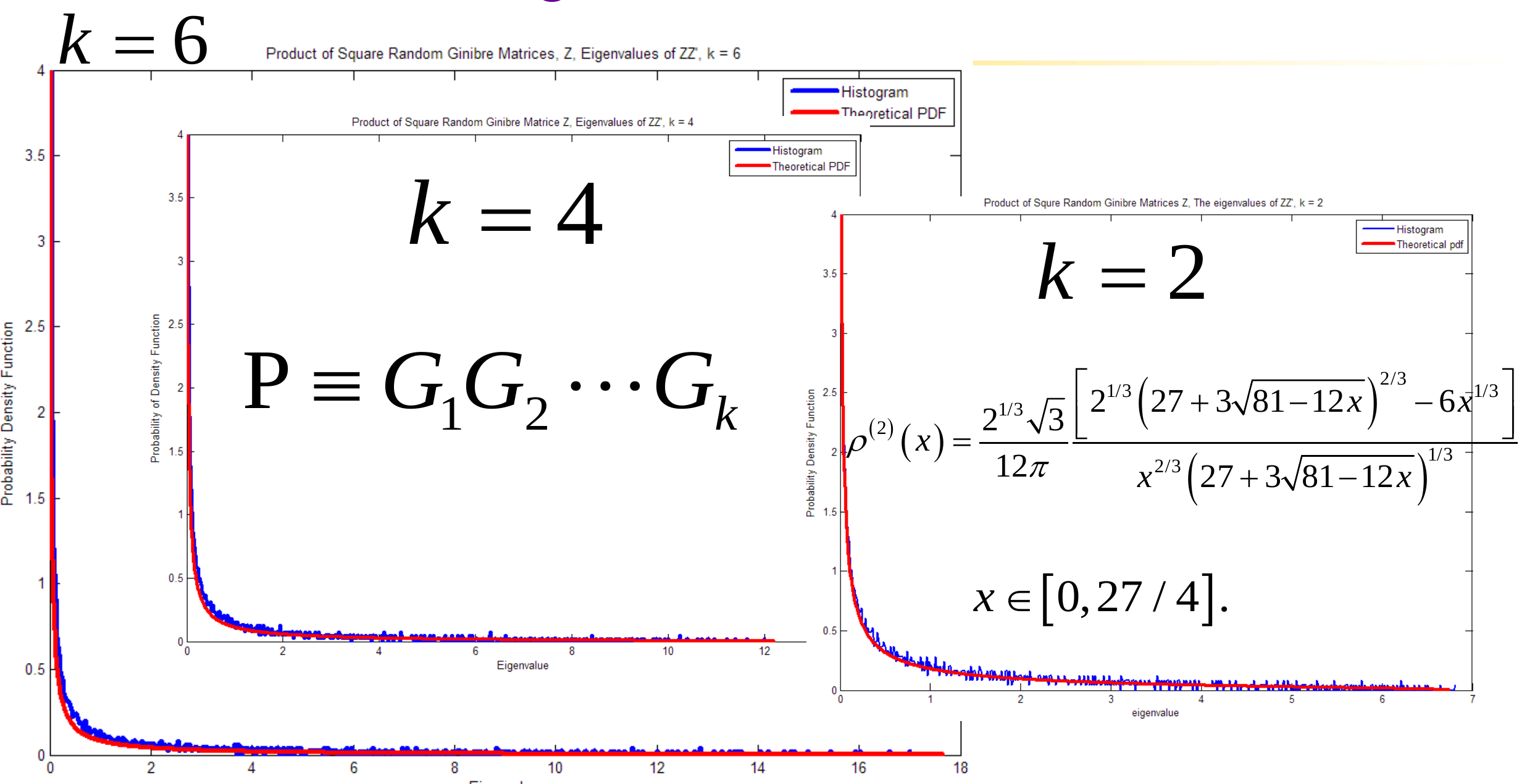}
	\caption{Product of Large Random Ginibre Matrices}
	\label{fig:ProductGinibre}
\end{figure}

Figure~\ref{fig:ProductGinibre} the product of large random Ginibre matrices. The product of $k$ large random Ginibre matrices has closed form. Gaussian is a special case of Ginibre matrix. 

\begin{figure}
	\centering
		\includegraphics[scale=0.6]{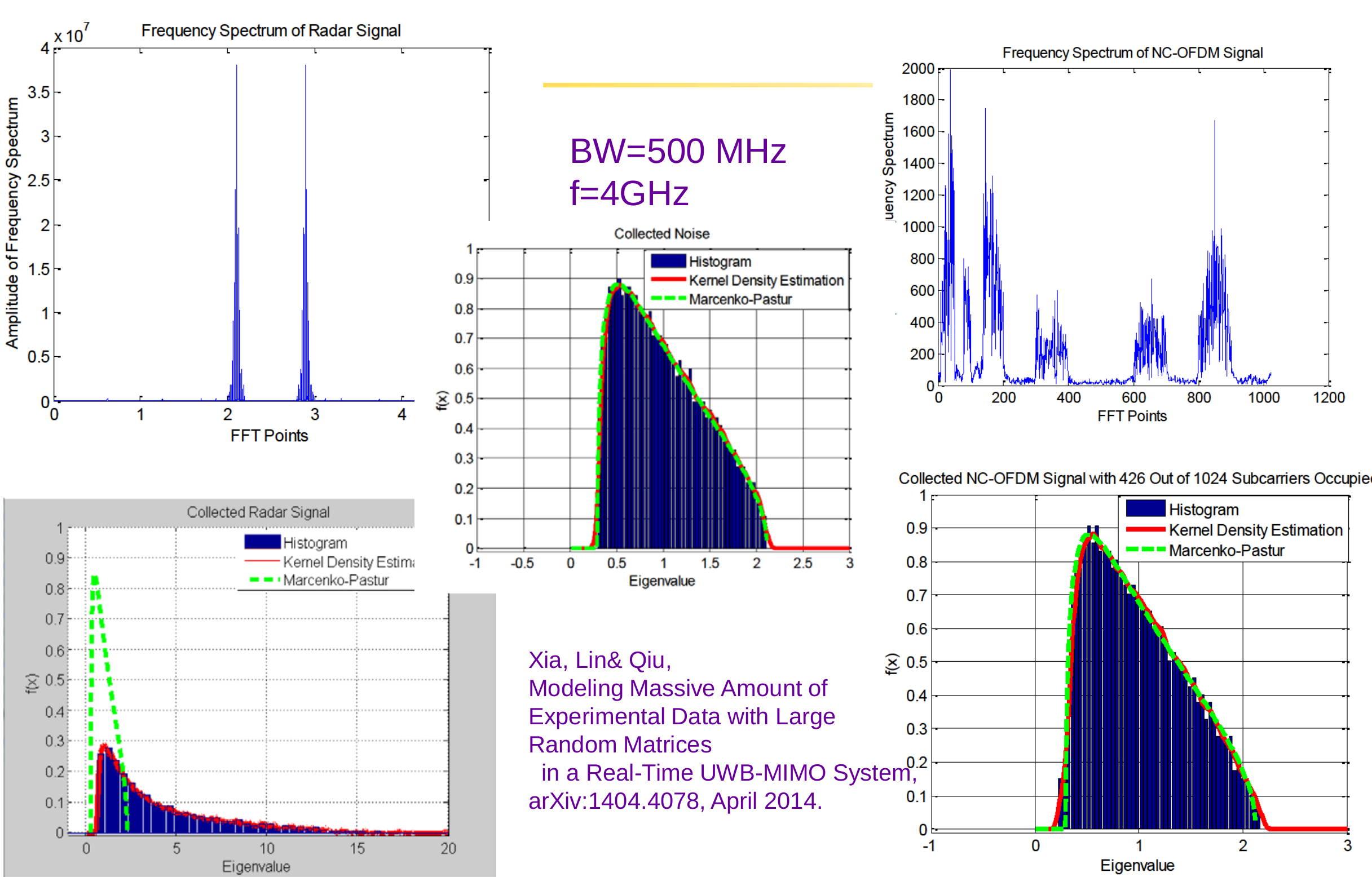}
	\caption{NC-OFDM Wideband Waveforms}
	\label{fig:NCOFDMwidebandwaveforms}
\end{figure}

Figure~\ref{fig:NCOFDMwidebandwaveforms} shows measured NC-OFDM wideband waveforms (bandwidth=500 MHz, central frequency=4GHz).  

\subsection{5G Massive MIMO: Euclidean Random Matrices}

In the next generation (5G) wireless system, massive MIMO uses hundreds (even thousands of) antennas that co-located or distributed in two dimensions (2D) or three dimensions (3D). 

\begin{figure}
	\centering
		\includegraphics[scale=0.6]{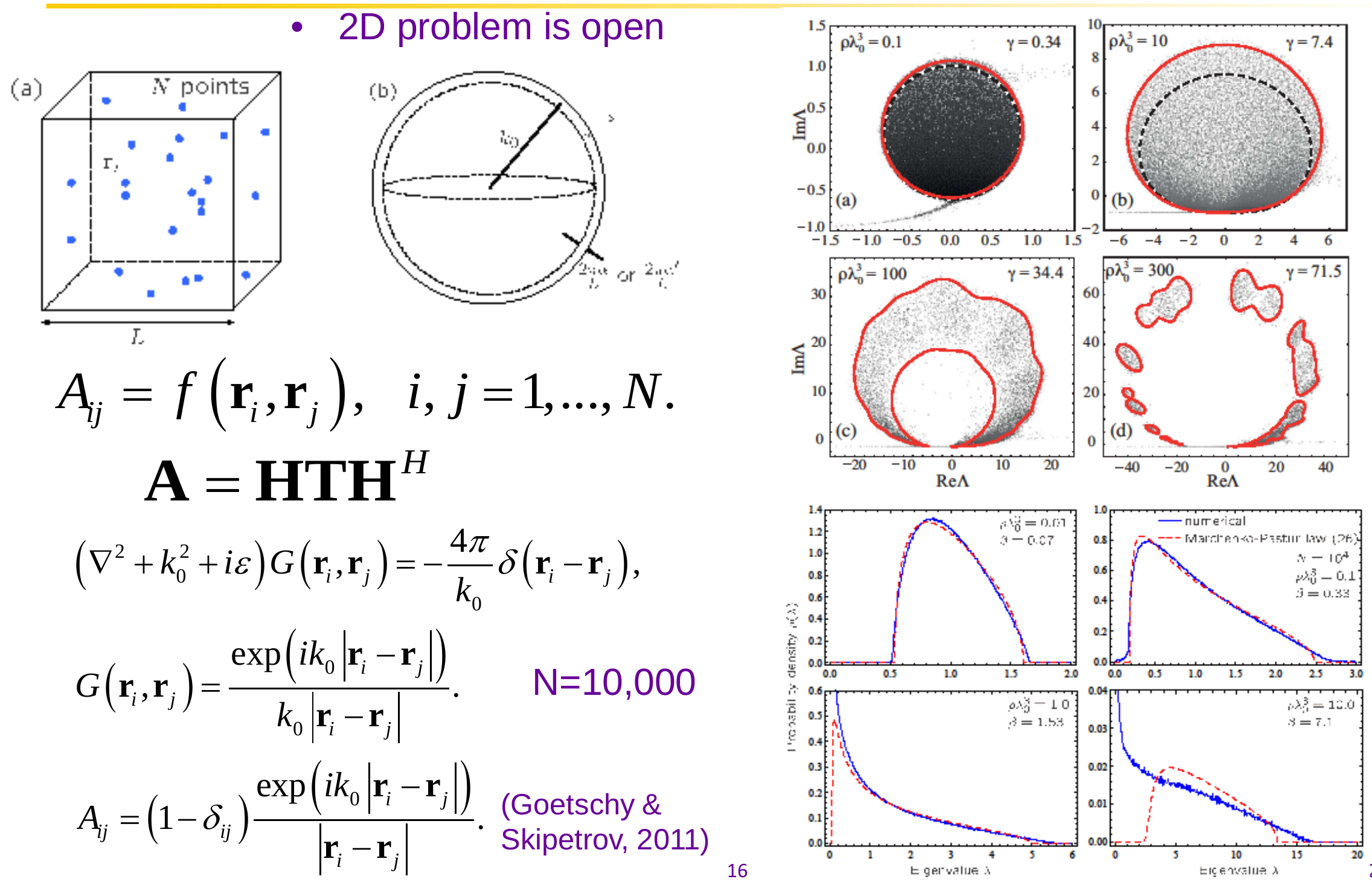}
	\caption{Massive MIMO: Euclidean Random Matrices}
	\label{fig:MassiveMIMOERM}
\end{figure}

Here we are interested in the massive datasets (random fields distributed in 3D). Modeling the random fields (or the channel) is a challenge. The idea here is to study the free-space propagation and thus the closed-form solution can be obtained~\cite{QiuAntonik2014Wiley}. Figure~\ref{fig:MassiveMIMOERM} shows massive MIMO modeled using Euclidean random matrices.

The Green's function between an arbitrary transmitter located at ${\bf r}_i$ and an arbitrary receiver located at ${\bf r}_j$ is expressed as
\begin{equation}
	\left( {{\nabla ^2} + k_0^2 + i\varepsilon_0 } \right)G\left( {{{\mathbf{r}}_i},{{\mathbf{r}}_j}} \right) =  - \frac{{4\pi }}{{{k_0}}}\delta \left( {{{\mathbf{r}}_i},{{\mathbf{r}}_j}} \right).
\end{equation}
where $k_0=2\pi/{\lambda _0}$ with $\lambda _0$ being the wavelength in free space, and $\varepsilon_0$ is the electric constant.  In free space, we obtain the closed-form expression \[G\left( {{{\mathbf{r}}_i},{{\mathbf{r}}_j}} \right) = \frac{{\exp \left( {i{k_0}\left| {{{\mathbf{r}}_i} - {{\mathbf{r}}_j}} \right|} \right)}}{{\left| {{{\mathbf{r}}_i} - {{\mathbf{r}}_j}} \right|}}.\]

The channel prorogation between an arbitrary transmitter located at ${\bf r}_i$ and an arbitrary receiver located at ${\bf r}_j$ is modeled by matrix entries
\begin{equation}
\label{GoetschySkipetrovmatrixentries}
	{A_{ij}} = f\left( {{{\mathbf{r}}_i},{{\mathbf{r}}_j}} \right),i,j = 1,...,N.
\end{equation}
Here $N=10,000$ is considered. It turns out that~\eqref{GoetschySkipetrovmatrixentries} can be represented in terms of the product of large random matrices
\begin{equation}
	{\mathbf{A}} = {\mathbf{HT}}{{\mathbf{H}}^H},
\end{equation}
leads to the use of free probability theory. In free space, we have \[{A_{ij}} = \left( {1 - {\delta _{ij}}} \right)\frac{{\exp \left( {i{k_0}\left| {{{\mathbf{r}}_i} - {{\mathbf{r}}_j}} \right|} \right)}}{{\left| {{{\mathbf{r}}_i} - {{\mathbf{r}}_j}} \right|}}.\]

The fundamental property of the large random matrix $\bf A$ is that $\bf A$ is non-Hermitian and thus its eigenvalues are complex in general, as shown in Figure~\ref{fig:MassiveMIMOERM}. Two free parameters determine the distribution of eigenvalues in the complex plane. One is the signal to-noise ratio $\gamma,$ and  the other is $\rho \lambda _0^3,$  and $\rho$ is the density of users (number of users per meter).

\section{A Non-Asymptotic Theory of Detection}
\label{sect:NonAympTheoryDet}

In~\cite{lin2012generalized}, we discover that the larger the length of the data, the bigger the gain (compared with the classical estimator-correlator test). It is commonly believed that the classical estimator-correlator test is optimal. To understand this effect, we are motivated for these three works~\cite{QiuBook2012Cognitive,Qiu_WicksBook2013,QiuAntonik2014Wiley}.

\begin{figure}
	\centering
		\includegraphics[scale=0.6]{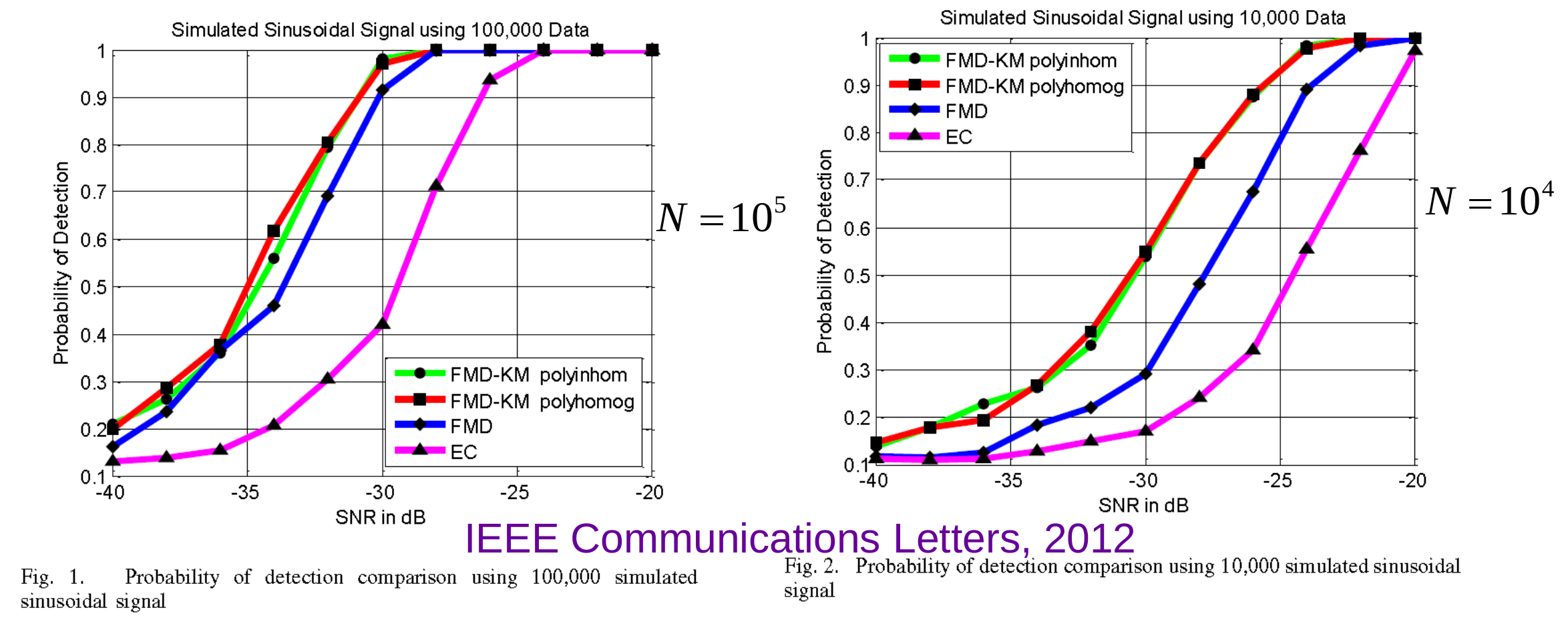}
	\caption{Detection Statistic with Functions of Random Matrices.}
	\label{fig:DetetcionStatistic}
\end{figure}

Let us give a tutorial introduction of this problem. We follow the classic Van Tree (1968)~\cite{VanTrees1968dem} for the classical treatment. For (column) random vectors \[{\mathbf{x}} = {\left[ {{X_1},...,{X_n}} \right]^T},{\mathbf{y}} = {\left[ {{Y_1},...,{Y_n}} \right]^T},{\mathbf{s}} = {\left[ {{S_1},...,{S_n}} \right]^T},\] we define their covariance matrices as \[{{\mathbf{R}}_x} = \mathbb{E}\left[ {{\mathbf{x}}{{\mathbf{x}}^T}} \right];{{\mathbf{R}}_y} = \mathbb{E}\left[ {{\mathbf{y}}{{\mathbf{y}}^T}} \right];{{\mathbf{R}}_x} = \mathbb{E}\left[ {{\mathbf{s}}{{\mathbf{s}}^T}} \right],\]
where $\mathbb E$ is the mathematical expectation. In practice, we only access $N$ independent copies of random vectors so form the random matrices \[{\mathbf{X}} = \left[ {{{\mathbf{x}}_1},...,{{\mathbf{x}}_N}} \right];{\mathbf{Y}} = \left[ {{{\mathbf{y}}_1},...,{{\mathbf{y}}_N}} \right];{\mathbf{S}} = \left[ {{{\mathbf{s}}_1},...,{{\mathbf{s}}_N}} \right].\] We are, in particular, interested in the following regime: $N$ and $n$ are large but finite! Now we are ready to compare three formulations 
\begin{equation}
\label{eq:comparethreeformulations}
	({\text{I}}){\text{ }}\left\{ \begin{gathered}
  {\mathcal{H}_0}:{\mathbf{y}} = {\mathbf{x}}, \hfill \\
  {\mathcal{H}_1}:{\mathbf{y}} = {\mathbf{s}} + {\mathbf{x}}, \hfill \\ 
\end{gathered}  \right.{\text{  }}\left( {{\text{II}}} \right){\text{ }}\left\{ \begin{gathered}
  {\mathcal{H}_0}:{{\mathbf{R}}_y} = {{\mathbf{R}}_x} \hfill \\
  {\mathcal{H}_1}:{{\mathbf{R}}_y} = {{\mathbf{R}}_s} + {{\mathbf{R}}_x} \hfill \\ 
\end{gathered}  \right.{\text{  }}\left( {{\text{III}}} \right){\text{ }}\left\{ \begin{gathered}
  {\mathcal{H}_0}:{\mathbf{Y}} = {\mathbf{X}} \hfill \\
  {\mathcal{H}_1}:{\mathbf{Y}} = {\mathbf{S}} + {\mathbf{X}}, \hfill \\ 
\end{gathered}  \right.
\end{equation}
where the first formulation follows from Van Tree (1968)~\cite{VanTrees1968dem}, and the second one is for quantum detection~\cite{QiuBook2012Cognitive}, and the third one is used in~\cite{lin2012generalized}. Formulation II is the limit form for 
\begin{equation}
\label{eq:classicalview}
	n{\text{ is fixed, but }}N \to \infty ,
\end{equation}
 which, however, is not true in general.   We are often interested in the non-asymptotic regime: 
\begin{equation}
\label{eq:modernview}
	N \to \infty ,n \to \infty {\text{  but  }}N/n \to c \in \left( {0,\infty } \right).
\end{equation}

Consider Formulation I first, for simplicity we assume that $\bf x$ a  zero-mean, Gaussian random vector. A vector $\bf z$ is a Gaussian random vector when its components $z_1,z_2,...,z_n$ are jointly Gaussian random variables. We further assume that \[\mathbb{E}\left[ {\mathbf{y}} \right] = \mathbb{E}\left[ {\mathbf{s}} \right] = {\mathbf{m}}.\] Of course $\bf m$ is a deterministic vector. In practice, we need to estimate $\bf m.$ Thus we have\[{{\mathbf{R}}_y} = \mathbb{E}\left[ {\left( {{\mathbf{y}} - {\mathbf{m}}} \right){{\left( {{\mathbf{y}} - {\mathbf{m}}} \right)}^T}} \right].\] The probability density of $\bf y$ is 
\[{p_{\mathbf{y}}} = \frac{1}{{{{\left( {2\pi } \right)}^{n/2}}{{\left( {\det {{\mathbf{R}}_y}} \right)}^{1/2}}}}\exp \left[ { - \frac{1}{2}{{\left( {{\mathbf{y}} - {\mathbf{m}}} \right)}^T}{\mathbf{R}}_{\mathbf{y}}^{ - 1}\left( {{\mathbf{y}} - {\mathbf{m}}} \right)} \right].\]
The likelihood ratio test follows easily: \[\Lambda  = \frac{{\frac{1}{{{{\left( {2\pi } \right)}^{n/2}}{{\left( {\det {{\mathbf{R}}_y}} \right)}^{1/2}}}}\exp \left[ { - \frac{1}{2}{{\mathbf{y}}^T}{\mathbf{R}}_{\mathbf{y}}^{ - 1}{\mathbf{y}}} \right]}}{{\frac{1}{{{{\left( {2\pi } \right)}^{n/2}}{{\left( {\det {{\mathbf{R}}_y}} \right)}^{1/2}}}}\exp \left[ { - \frac{1}{2}{{\left( {{\mathbf{y}} - {\mathbf{m}}} \right)}^T}{\mathbf{R}}_{\mathbf{y}}^{ - 1}\left( {{\mathbf{y}} - {\mathbf{m}}} \right)} \right]}}\begin{array}{*{20}{c}}
   {\mathop  > \limits^{{\mathcal{H}_1}} }  \\ 
   {\mathop  < \limits_{{\mathcal{H}_0}} }  \\ 
\end{array} \eta .\]
Taking logarithms, we obtain 
\begin{equation}
\label{eq:LRTlog}
	\frac{1}{2}{\left( {{\mathbf{y}} - {\mathbf{m}}} \right)^T}{\mathbf{R}}_{\mathbf{y}}^{ - 1}\left( {{\mathbf{y}} - {\mathbf{m}}} \right) - \frac{1}{2}{{\mathbf{y}}^T}{\mathbf{R}}_{\mathbf{y}}^{ - 1}{\mathbf{y}}\begin{array}{*{20}{c}}
   {\mathop  > \limits^{{\mathcal{H}_1}} }  \\ 
   {\mathop  < \limits_{{\mathcal{H}_0}} }  \\ 
\end{array} \ln \Lambda  + \frac{1}{2}\ln \det {{\mathbf{R}}_y} - \frac{1}{2}\ln \det {{\mathbf{R}}_x} \triangleq {\gamma ^ \star }.
\end{equation}

Consider the special case: 
\begin{equation}
\label{eq:whiteGaussiancov}
	{{\mathbf{R}}_y} = {{\mathbf{R}}_{\mathbf{x}}}={\bf R}.
\end{equation} 
Equation~\eqref{eq:LRTlog} is equivalent to \[l\left( {\mathbf{y}} \right) = {{\mathbf{m}}^T}{\mathbf{Ry}}\begin{array}{*{20}{c}}
   {\mathop  > \limits^{{\mathcal{H}_1}} }  \\ 
   {\mathop  < \limits_{{\mathcal{H}_0}} }  \\ 
\end{array} {\gamma _*}.\]
The quantity on the left is a scalar Gaussian random variable, for it was obtained by a linear transformation of jointly Gaussian random variables. 
The distance between the means on the two hypotheses is defined as
\begin{equation}
	{d^2} \triangleq \frac{{{{\left[ {\mathbb{E}\left( {l|{\mathcal{H}_1}} \right) - \mathbb{E}\left( {l|{\mathcal{H}_1}} \right)} \right]}^2}}}{{\operatorname{Var} \left( {l|{\mathcal{H}_1}} \right)}}.
\end{equation}
After some manipulation, we obtain 
\begin{equation}
\label{eq:distancettt}
	{d^2} = {{\mathbf{m}}^T}{\mathbf{Rm}}.
\end{equation}
When for white Gaussian noise 
\begin{equation}
\label{eq:Gaussianwhitecovr} 
	{\mathbf{R}} = {\sigma ^2}{{\mathbf{I}}_n},
\end{equation}
 we have
\begin{equation}
\label{eq:distancetttGaussian}
	{d^2} = \frac{{{{\mathbf{m}}^T}{\mathbf{m}}}}{{{\sigma ^2}}} = \frac{{{{\left\| {\mathbf{m}} \right\|}^2}}}{{{\sigma ^2}}}
\end{equation}

where $||\cdot||$ is the Euclidean norm of a vector. ${{{\left\| {\mathbf{m}} \right\|}^2}}$ can be viewed as the signal power.

Now let us turn our attention Formulation III in~\eqref{eq:comparethreeformulations}. We can always write
\begin{equation}
\label{eq:FormulationIIIDetection}
	\begin{gathered}
  {\mathcal{H}_0}:\frac{1}{N}{{\mathbf{Y}}^H}{\mathbf{Y}} = \frac{1}{N}{{\mathbf{X}}^H}{\mathbf{X}} = \frac{1}{N}\sum\limits_{i = 1}^N {{\mathbf{x}}_i^H{{\mathbf{x}}_i}} , \hfill \\
  {\mathcal{H}_1}:\frac{1}{N}{{\mathbf{Y}}^H}{\mathbf{Y}} = \frac{1}{N}{{\mathbf{S}}^H}{\mathbf{S}} + \frac{1}{N}{{\mathbf{X}}^H}{\mathbf{X}} = \frac{1}{N}\sum\limits_{i = 1}^N {{\mathbf{s}}_i^H{{\mathbf{s}}_i}}  + \frac{1}{N}\sum\limits_{i = 1}^N {{\mathbf{x}}_i^H{{\mathbf{x}}_i},}  \hfill \\ 
\end{gathered} 
\end{equation}
where ${{{\mathbf{x}}_1},...,{{\mathbf{x}}_N}}$ are $N$ independent copies of the random vector $\bf x$ and ${{{\mathbf{s}}_1},...,{{\mathbf{s}}_N}}$ are $N$ independent copies of the random vector ${\bf s}.$ 

In this modern non-asymptotic view of~\eqref{eq:modernview}, the derivation of Formulation I is not justified. The central argument is the expectation operation ${\mathbb E}[\cdot]$ in analysis, which requires~\eqref{eq:classicalview}, indirectly through the central limit theorem~\cite{jacod2003probability}. 

\begin{theorem}[Central Limit Theorem] Let ${\left( {{{\mathbf{x}}_i}} \right)_{i \geqslant 1}}$ be i.i.d. ${\mathbb{R}^n}-$valued random variables. Let the (vector) $\boldsymbol\mu  = \mathbb{E}\left[ {{{\mathbf{x}}_i}} \right],$ and let ${\mathbf{\Sigma }}$ denote the covariance matrix: ${\mathbf{\Sigma }} = {\left( {{\Sigma _{k,l}}} \right)_{1 \leqslant k,l \leqslant n}},$ where ${\Sigma _{k,l}} = \operatorname{Cov} \left( {X_i^k,X_i^l} \right),$ where ${X_i^k}$ is the $k-$th component of the ${\mathbb{R}^n}-$valued random variable ${\bf x}_i.$ Then\[\mathop {\lim }\limits_{N \to 0} \frac{{{{\mathbf{s}}_N} - N\mu }}{{\sqrt N }} = {\mathbf{z}},\]
where ${{\mathbf{s}}_N} = \sum\limits_{i = 1}^N {{{\mathbf{x}}_i}} $ and the distribution of $\bf z$ is zero-mean Gaussian $\mathcal{N}\left( {{\mathbf{0}},{\mathbf{\Sigma }}} \right)$ and where the convergence is in distribution.
\end{theorem}
The central limit theorem is not valid for non-asymptotic regime~\eqref{eq:modernview}.
 
\begin{figure}
	\centering
		\includegraphics[scale=0.6]{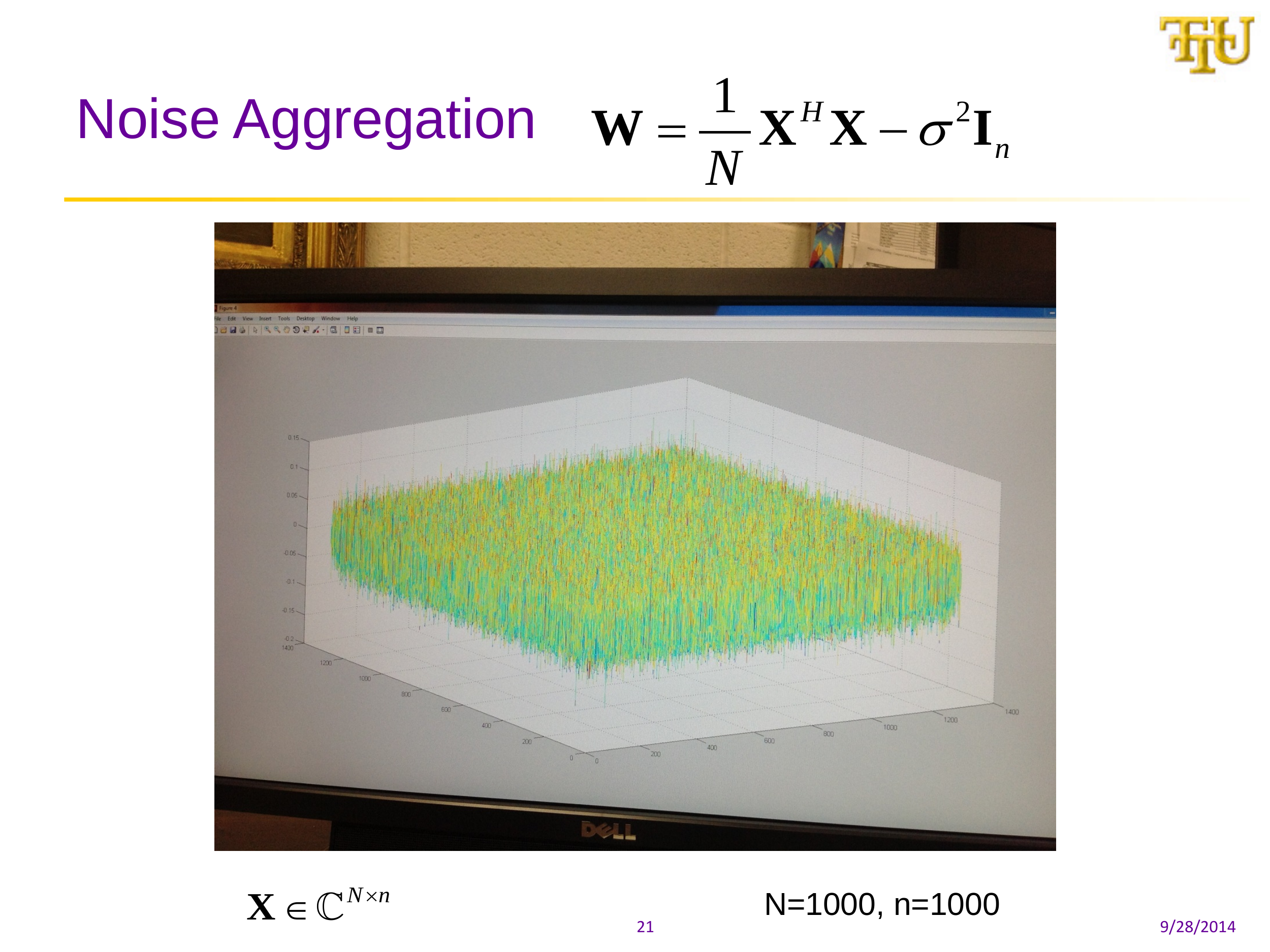}
	\caption{Noise aggregation ${\bf W}_n$ defined in~\eqref{eq:hollowWishartmatrix }. N=1,000, n=1,000.}
	\label{fig:holoowWishart}
\end{figure}

Consider an example for white Gaussian noise,~\eqref{eq:Gaussianwhitecovr} cannot be justified. We consider the hollow Wishart matrix 
\begin{equation}
\label{eq:hollowWishartmatrix }
	{{\mathbf{W}}_n} = \frac{1}{N}\sum\limits_{i = 1}^N {{\mathbf{x}}_i^H{{\mathbf{x}}_i}}  - {\sigma ^2}{{\mathbf{I}}_n}.
\end{equation}
For massive data, the noise is aggregated. The effect of noise aggregation cannot be ignored, as shown in Figure~\ref{fig:holoowWishart}. Understanding noise aggregation is central to big data problems.
Without loss of generality, Formulation III~\eqref{eq:FormulationIIIDetection} can be rewritten as 
\begin{equation}
\label{eq:FormulationIIIDetectionhollow}
	\begin{gathered}
  {\mathcal{H}_0}:{{\mathbf{W}}_n}, \hfill \\
  {\mathcal{H}_1}:\frac{1}{N}\sum\limits_{i = 1}^N {{\mathbf{s}}_i^H{{\mathbf{s}}_i}}  + {{\mathbf{W}}_n} \hfill \\ 
\end{gathered} 
\end{equation}
where ${\bf W}_n$ is the $n\times n$ random matrix. Our intuition is that the trace (sometimes called average) of ${\bf W}_n$ tends to vanish when $n$ is large enough. The central task is to find the non-asymptotic relation as a function of $n.$

Using concentration inequalities, it is shown in~\cite{Qiu_WicksBook2013} that the scalar random variable 
\begin{equation}
\label{eq:ZtraceWn} 
	Z = \operatorname{Tr} \left( {{{\mathbf{W}}_n}} \right)
\end{equation}
behaves very nicely. We obtain~\cite[p.221]{Qiu_WicksBook2013} 

\begin{equation}
\label{eq:ZtraceWnVar} 
	\mathbb{E}\left[ Z \right] = 0,\;\;\;{\kern 1pt} \operatorname{Var} \left[ Z \right] \leqslant \frac{c}{{{n^2}}},
\end{equation}
where $c$ is a constant. We guess that the convergence rate $1/n^2$ may be optimal. Our algorithm claims ${\mathcal H}_1$ when $Z$ is above zero. 

\section{The Mathematical Foundation of Big Data}
In the next subsection, we take material from the NSF workshop~\cite{Tian2013bigdata}.
\subsection{Overview of Big Data}

\begin{itemize}
	\item Big engineering data has unique characteristics: more disciplined and regulated.
	\item  Emerging engineering systems with big-data opportunities:  smart grids, sensor nets, transportation, telemedicine, aerospace, testing, safety, nuclear, design blueprints and more.
	\item Rethink of data collection and storage to facilitate big data processing and inference tasks.
	\item How do we trade-off complexity for accuracy in massive decentralized signal and data analysis tasks?
	\item How can efficient signal and data analysis algorithms be developed for big, unstructured or loosely structured data?
	\item What are the basic principles and useful methodologies to scale inference and learning algorithms and trade off  the computational resources (e.g., time, space and energy) according to the needs of engineering practice (e.g. robustness vs. efficiency, real-time)?

\end{itemize}
Big Data processing and analysis is summarized below 
\begin{itemize}
	\item Integration of very heterogeneous data: correlation mining in massive database; 
		\item Data at vastly different scales and noise levels; Mixture of continuous and categorical variables.
	\item Reliable and robust quantitative models: Uncertainty quantitation; Adaptive to drift over time.
	\item High throughput real-time processing: Smart adaptive sampling and compression; Distributed or parallel processing architectures.
	\item Interactive user interfaces: Human-in-the-loop processing;
	\item Visualization and dimensionality reduction.
\end{itemize}
Some signal processing challenges include
\begin{itemize}
	\item Heterogeneous data integration: Ranking signals for human-aided selection of relevant variables; Fusing graphs, tensors, and sequence data; Active visualization: dimensionality reduction.
\item Flexible low complexity modeling and computation: Scalable signal processing: distributed algorithms and implementation; Smart sampling: feedback controlled signal search and acquisition.
\item Reliable robust models for anomaly detection and classification: Parsimonious signal processing: Sparse correlation graphical models; Decomposable signal processing: factored models and algorithms
\end{itemize}

Large random matrices for Big Data. Random matrices play a central role in statistics in the context of multivariate data. The continued growth of big data has given rise to high dimensional statistics analysis. Convex analysis, Riemannian geometry and combinatorics are relevant.
 Random matrix theory (RMT) has emerged as a particularly useful framework for many theoretical questions associated with the analysis of high-dimensional multivariate data.
RMT affects the modern statistical thinking in two ways. Most of the mathematical treatments of RMT have focused on matrices with high degree of independence in the entries, which one may refer to as ``unstructured" random matrices. Recall that about 75$\%$ of big data are unstructured.
 In high-dimensional statistics, we are primarily interested in problems where there are lower dimensional structures buried under random noise.

Traditional statistical theory, particularly in multivariate analysis, did not contemplate the demands of high dimensionality in data analysis. Classical multivariate analysis textbooks consider: (1)  Only asymptotically optimum, not optimum for any size data; (2) dimension of the dataset, $p$  is fixed small constant or at least negligible compared with the sample size  $n.$
Modern datasets have: (1) Their dimensions  $p$  can be proportionally large compared with the sample size $n.$ (2) Examples: sensor network data,  wireless network data, smart grid data, financial data, consumer data, manufacturing data, multimedia data, etc. (3) The likelihood ratio test (LRT), the classical statistical inference, must be revisited.

In this next subsection, we draw material heavily from~\cite{vanprobability}. 
\subsection{Probability in High Dimension}
``High dimension'' refers to the presence of many distinct but interacting random variables, including
\begin{itemize}
		\item Large random structure: random matrices, random graphs,…
	\item Statistics and machine learning: estimation, prediction and model selection for high-dimensional data.
	\item Randomized algorithms in computer science.
	\item Random codes in information theory.
	\item Statistical physics: Gibbs measure, percolation, spin glasses,…
	\item Random combinatorial structures: longest increasing subsequence, spanning
trees, travelling salesman problem, . . .
	\item Probability in Banach spaces: probabilistic limit theorems for Banach valued random variables, empirical processes, local theory of Banach spaces, geometric functional analysis, convex geometry.
 	\item Mixing times and other phenomena in high-dimensional Markov chains. 
\end{itemize}

Non-asymptotic probabilistic inequalities are especially relevant when the number of observables is large. They are suitable for large complex sets of data with possibly huge collections of models at different scales. This framework allows the collection of models together with their dimensions $n$ to vary freely, letting the dimensions $n$ be possibly of the same order of magnitude as the number of observations $p.$  We are not concerned with limit theorems (as in many classical probabilistic results), but rather with explicit estimates that are either dimension-free, or that capture precisely the dependence of the problem on the relevant dimensional parameters. In \textit{asymptotic} results, one must take all these parameters to the limit in a fixed relation to one another, while the \textit{non-asymptotic} viewpoint allows to express the interrelation between the different parameters in a much more precise way.

High-dimensional problems typically involve interactions between a large number of degrees of freedom whose aggregate contributions to the phenomenon of interest must be accounted for in the mathematical analysis;  The explicit nature of non-asymptotic estimates makes them particularly well suited to be used as basic ingredients of the analysis, even if the ultimate result of interest is asymptotic in nature.  

In Section~\ref{sect:NonAympTheoryDet}, we give an introdtuction to the non-asymptotic theory of  random matrix-valued hypothesis detection. In 2010,  we were lucky to discover that concentration inequalities were indeed the probabilistic tools.

Three basic principles formulated informally as ``theorems'' are pointed out below. Of course, we need to make precise some terms. But rigor is not concern here.

\begin{theorem} [Concentration]
If ${X_1},...,{X_n}$ are independent (or weakly dependent) random variables,
then the random variable  $f\left( {{X_1},...,{X_n}} \right)$ is ``close'' to its mean $\mathbb{E}\left[ {f\left( {{X_1},...,{X_n}} \right)} \right],$ provided that the function  $f\left( {{x_1},...,{x_n}} \right)$ 
is not too ``sensitive'' to any of the coordinates $x_i,i=1,...,n.$ 
\end{theorem}

This theorem is valid for very complicate functions such as eigenvalues of large random matrices. Sometimes the expectation of the function $\mathbb{E}\left[ {f\left( {{X_1},...,{X_n}} \right)} \right]$may be unknown. Even in this scenario, concentration inequalities can still be used to find the fluctuations of this function.  When the Central Limit Theorem is invalid, the next best thing you can do is to resort to concentration inequalities.

\begin{theorem} [Supremum]
If the random process  ${\left\{ {{X_t}} \right\}_{t \in T}}$  is ``sufficiently continuous'', then the magnitude of the supremum $\mathop {\sup }\limits_{t \in T} {X_t}$  is controlled by the ``complexity'' of the index set $T.$
\end{theorem}
A family of random variables indexed by a set $T$ that is frequently high- or infinite-dimensional.
We often involve a large number of interdependent degrees of freedom; there is a need to obtain \textit{simultaneous control} over many random variables. The investigation of suprema is in fact surprisingly general. Four functions are defined below in this framework. The first one is the operator norm of a matrix $\bf A,$ \[\left\| {\mathbf{A}} \right\| = \mathop {\sup }\limits_{{\mathbf{u}},{\mathbf{v}} \in {B_2}} \left\langle {{\mathbf{u}},{\mathbf{Av}}} \right\rangle ,\]
where $\left\langle \cdot \right\rangle $ represents the inner product of two vectors, $B_2$ is the Euclidean unit ball in ${\mathbb R}^n.$  

The second function is related to the norms of random vectors. Let $X$ be a random vector in ${\mathbb R}^n,$ and let ${\left\| \cdot \right\|_B}$ be any norm on ${\mathbb R}^n,$where $B$ denotes the unit ball of ${\left\| \cdot \right\|_B}.$ The duality theory of Banach spaces implies that we can write \[{\left\| X \right\|_B} = \mathop {\sup }\limits_{t \in {B^o}} \left\langle {t,X} \right\rangle ,\]
where ${B^ \circ }$ denotes the dual ball. The supremum of the random processes \[{\left\{ {{X_t} = \left\langle {t,X} \right\rangle } \right\}_{t \in {B^o}}}\] arises naturally in probability in Banach spaces.

The third function is related to empirical risk minimization. The problem is to estimate how close the empirical risk minimizer is to the optimal parameter as a function of the number of samples $n,$ the dimension of the parameter space $\Theta ,$ the dimension of the state space $\bf X,$ etc. Many problems in statistics and machine learning may be formulated as the problem of computing \[\mathop {\arg \min }\limits_{\theta  \in \Theta } \mathbb{E}\left[ {l\left( {\theta ,X} \right)} \right]\]
given only observed ``data'' consisting of i.i.d. samples ${X_1},...,{X_n} \sim X.$ That is to, without knowledge of the law of $X$. Here $l$F is a given loss function, and $\Theta$ is a given parameter space, which depends on the problem at hand. Formally we have
\[\mathop {\sup }\limits_{\theta  \in \Theta } \frac{1}{n}\sum\limits_{i = 1}^n {\left\{ {l\left( {\theta ,{X_i}} \right) - \mathbb{E}\left[ {l\left( {\theta ,{X_i}} \right)} \right]} \right\}},\] which is the supremum of a random process. 

The fourth function is related a convex function $f:{\mathbb{R}^n} \to \mathbb{R}$ \[f\left( {\mathbf{x}} \right) = \mathop {\sup }\limits_{{\mathbf{y}} \in {\mathbb{R}^n}} \left\{ {\left\langle {{\mathbf{y}},{\mathbf{x}}} \right\rangle  - {f^*}\left( {\mathbf{y}} \right)} \right\},\] where ${{f^*}}$ denotes the convex conjugate of $f.$

Now we are ready to talk about the third principle: universality. The high-dimensional phenomenon is, in a sense, robust to the precise details of the model ingredients.

\begin{theorem}[Universality]
If   ${X_1},...,{X_n}$ are independent (or weakly dependent) random variables,  then the expectation  $\mathbb{E}\left[ {f\left( {{X_1},...,{X_n}} \right)} \right]$ is ``insensitive'' to the distribution of   when the function $f$ is ``sufficiently smooth''.
\end{theorem}

 Let $X_1,X_2,...,$ be i.i.d. random variables with finite variance.  The law of large numbers says \[\frac{1}{{\sqrt n }}\sum\limits_{i = 1}^n {\left\{ {{X_i} - \mathbb{E}{X_i}} \right\}}  \to 0{\text{   as  }}n \to \infty .\] We do not only know that the fluctuations are of
order ${n^{ - 1/2}}$ (as is captured by the concentration phenomenon), but we have
much more precise information as well: by the central limit theorem, we have
a precise description of the distribution of the fluctuations, as \[\frac{1}{{\sqrt n }}\sum\limits_{i = 1}^n {\left\{ {{X_i} - \mathbb{E}{X_i}} \right\}}  \approx {\text{Gaussian}}\]
when $n$ is large. A different way of saying this property is that
\[\frac{1}{{\sqrt n }}\sum\limits_{i = 1}^n {\left\{ {{X_i} - \mathbb{E}{X_i}} \right\}}  \approx \frac{1}{{\sqrt n }}\sum\limits_{i = 1}^n {\left\{ {{G_i} - \mathbb{E}{\mathbb{G}_i}} \right\}} ,\]
where $G_k$ are independent Gaussian random variables with the same mean
and variance of $X_k$ (here $ \approx$ denotes closeness of the distributions). 

As in the case of concentration, it turns out that this phenomenon is not
restricted to linear functions of independent random variables, but is in fact
a manifestation of a more general principle.

\subsection{Talagrand's Concentration of Measure }
\subsection{Free Entropy}

\section{Conclusion}
Big data problems are ubiquitous. The basic questions: How do we represent the massive datasets? How do we make sense of those data? We promote the non-asymptotic statistical viewpoint, that is built upon concentration inequalities. Data-driven system design is enabled. Large Random matrices serve as basic building blocks. We only know little about the Big Data paradigm. One wonder if new Data Science is emerging for today's and future's  challenges. 

\section*{ACKLOWLEDGEMENT}
 
   This exposition paper  is based on two sources: ``The Foundation of Big Data: Experiments, Formulation, and Applications'', 4 invited talks in the summer of 2014, at State Grid Research Institute, Shanghai JiaoTong University, University of Electronic Science and Technology of China, Jiling Univerisity, and an more recent invited talk entitled ``Big Data of a Large-Scale Cognitive Radio Network: Testbed, Data Representation and Analytics'', at Wireless Networks for Big Data, Hefei, Anhui, China, September 29-30, 2014. This exposition paper was also used as lecture notes in ECE 7750 in TTU for PhD students. 
	
	The author thanks Guangrong Yue for his outstanding presentation of this work at Hefei.  The author also wants to thank Shaoqian Li for useful discussions and hospitality. He also wants to thank his team, especially and Changchun (Alex) Zhang, Nan (Terry) Guo, Zhen (Edward) Hu, Feng Lin, and Shujie Hou, for their contributions that make this paper possible. Last but not least, he want to thank the following for their hospitality:  Xiucheng Jiang, Zejian Jin, Shaoqian Li,  Xiaoyi Sun,  Jusheng Yu, Wenxian Yu,  Guangrong Yue,   Dongxia Zhang, and Chaoyang Zhu.
	
	This work is in part funded by National Science Foundation through three grants (ECCS-0901420, ECCS-0821658, and CNS-1247778), and Office of Naval Research through two grants (N00010-10-1-0810 and N00014-11-1-0006).


    \bibliographystyle{ieeetr}   

 \begin {small}
 
 \bibliography{Bible/5GWirelessSystem,Bible/Big_Data,Bible/Compressed_Sensing,Bible/Smart_Grid,Bible/Graph_Complex_Network,Bible/Machine_Learning,Bible/Convex_Optimization,Bible/LowRankMatrixRecovery,Bible/Concentration_of_Measure,Bible/ClassicalMatrixInequalities,Bible/CompeltelyPositiveMaps,Bible/QuantumChannel,Bible/QuantumHypothesisTesting,Bible/TraceInequalities,Bible/QuantumInformation,Bible/Matrix_Inequality,Bible/RandomMatrixTheory,Bible/Fractional_Integration_bib,Bible/UWB_bib,Bible/Qiu_Group_bib,Bible/LTI_Comm_Theory_bib,Bible/Software_Defined_Radio_bib,Bible/Time_Reversal_bib,Bible/MIMO_bib,Bible/Radar_Waveform_Optim_bib,Bible/Information_Theory_bib,Bible/Compressed_Sensing_Theory,Bible/Compressed_Sensing_UWB,Bible/MISC_bib,Bible/CS_Applications,Bible/CS_Data_Stream_Algorithms,Bible/CS_Extensions,Bible/CS_Foundations_Connections,Bible/CS_Multisensor_Distributed,Bible/CS_Recovery_Algorithms,Bible/CognitiveRadio/Cognitive_Radio_bib,Bible/CognitiveRadio/CognitiveRadio2008_bib,Bible/CognitiveRadio/DySpan2007,Bible/CognitiveRadio/DySpan2005,Bible/CognitiveRadio/Gardner,Bible/CognitiveRadio/jsac200703,Bible/CognitiveRadio/jsac200801,Bible/CognitiveRadio/sensingDTV,Bible/CognitiveRadio/ucberkeley,Bible/CognitiveRadio/UWBCognitiveRadio_bib,Bible/CognitiveRadio/IEEE_JSSP_2008,Bible/CognitiveRadio/Bayesian_Network_Cognitive_Radio,Bible/CognitiveRadio/Exploiting_Historical_Spectrum,Bible/CognitiveRadio/Duke_Carin,Bible/CognitiveRadio/LiHu_upper,Bible/CognitiveRadio/AFRLref,Bible/CognitiveRadio/SVM,Bible/CognitiveRadio/NSF_ECCS_0821658,Bible/CognitiveRadio/SmartGrid}

\end {small}
	
 \end{document}